\documentclass[twocolumn,superscriptaddress,showpacs,preprintnumbers,amsmath,amssymb]{revtex4}

\usepackage{graphicx}
\usepackage{dcolumn}
\usepackage{bm}

\begin{document}

\preprint{}

\preprint{}
\title{Imaging of current induced N{\'e}el vector switching in antiferromagnetic Mn$_2$Au}

\author{S.\,Yu.\,Bodnar}
\affiliation{Institute of Physics, Johannes Gutenberg-University, 55099 Mainz, Germany}
\author{M. Filianina}
\affiliation{Institute of Physics, Johannes Gutenberg-University, 55099 Mainz, Germany}
\affiliation{Graduate School Materials Science in Mainz, Staudingerweg 9, 55128 Mainz, Germany}
\author{S.\,P.\, Bommanaboyena}
\affiliation{Institute of Physics, Johannes Gutenberg-University, 55099 Mainz, Germany}
\author{T. Forrest}
\affiliation{Diamond Light Source, Chilton, Didcot, Oxfordshire, OX11 0DE, United Kingdom}
\author{F. Maccherozzi}
\affiliation{Diamond Light Source, Chilton, Didcot, Oxfordshire, OX11 0DE, United Kingdom}
\author{A.\,A. Sapozhnik}
\affiliation{Institute of Physics, Johannes Gutenberg-University, 55099 Mainz, Germany}
\affiliation{Graduate School Materials Science in Mainz, Staudingerweg 9, 55128 Mainz, Germany}
\author{Y. Skourski}
\affiliation{HLD-EMFL, Helmholtz-Zentrum Dresden-Rossendorf, 01328 Dresden, Germany }
\author{M. Kl\"aui}
\affiliation{Institute of Physics, Johannes Gutenberg-University, 55099 Mainz, Germany}
\affiliation{Graduate School Materials Science in Mainz, Staudingerweg 9, 55128 Mainz, Germany}
\author{M. Jourdan}
\affiliation{Institute of Physics, Johannes Gutenberg-University, 55099 Mainz, Germany}
\affiliation{Graduate School Materials Science in Mainz, Staudingerweg 9, 55128 Mainz, Germany}
\email{Jourdan@uni-mainz.de}

\date{\today}

\begin{abstract}
The effects of current induced N{\'e}el spin-orbit torques on the antiferromagnetic domain structure of epitaxial Mn$_2$Au thin films were investigated by X-ray magnetic linear dichroism - photoemission electron microscopy (XMLD-PEEM). We observed current induced switching of AFM domains essentially corresponding to morphological features of the samples. Reversible as well as irreversible N{\'e}el vector reorientation was obtained in different parts of the samples and the switching of up to 30~\% of all domains in the field of view of 10~$\mu$m is demonstrated. Our direct microscopical observations are compared to and fully consistent with anisotropic magnetoresistance effects previously attributed to current induced N{\'e}el vector switching in Mn$_2$Au.
\end{abstract}

\maketitle

In antiferromagnetic (AFM) spintronics the staggered magnetization, or more precisely the N{\'e}el vector describing the spin structure, can be used to encode information \cite{Jun18,Bal18,Jung18}. For the switching of the N{\'e}el vector and the read-out of its orientation different strategies have been pursued \cite{Son18}. The N{\'e}el vector was e.\,g.\,manipulated by an exchange-spring effect with a ferromagnet (FM)  and read-out via tunneling anisotropic magnetoresistance (T-AMR) measurements \cite{Par11, Fin14}. Other experiments were based on a ferromagnet to AFM phase transition \cite{Mar14} or on strain induced anisotropy modifications \cite{Sap17}. However, for antiferromagnetic spintronics N{\'e}el vector switching by current-induced spin-orbit torques (SOTs) \cite{Zel18}, whose FM counterparts are already established for memory applications \cite{Gam11,Bra14}, are most promising due to superior scaling, switching speed and device compatibility.

The SOTs used for FM spintronics are typically generated at interfaces with heavy metals \cite{Mir11, Liu12}. However, a specific crystallographic structure with oppositely broken inversion symmetry on the each of the collinear AFM sublattices makes Mn$_2$Au and CuMnAs up to now the only known antiferromagnets, for which a so called bulk N{\'e}el spin-orbit torque (NSOT) \cite{Zel14} can enable current induced N{\'e}el vector manipulation in a single layer system. Indeed, this was demonstrated  experimentally for CuMnAs \cite{Wad16, Wad18} and, more recently, for Mn$_2$Au \cite{Bod18, Zho18, Mei18} as well.

Whereas in the case of CuMnAs, the modification of the AFM domain structure by current pulses was observed directly by X-ray magnetic linear dichroism - photoelemission electron microscopy (XMLD-PEEM) \cite{Wad18, Grz17}, such microscopic insights are missing for Mn$_2$Au up to now. However, direct imaging of the effect of current pulses on the N{\'e}el vector orientation is crucial for the interpretation of previously published results of resistivity changes attributed to a N{\'e}el vector reorientation in Mn$_2$Au \cite{Bod18, Zho18, Mei18}. Furthermore, magnetic microscopy enables the identification of important quantities and mechanisms of the N{\'e}el vector manipulation such as switched volume fraction, morphological influence on the domain pattern, and domain wall motion.

In this paper we demonstrate the imaging of current induced modifications of the AFM domain structure of epitaxial Mn$_2$Au thin films and relate the results to previous transport measurements.

We investigated epitaxial Mn$_2$Au(001) thin films deposited by RF sputtering from a single stoichiometric target directly onto heated Al$_2$O$_3$$(1\bar{1}02)$ substrates. Although there is no obvious epitaxial relation of Mn$_2$Au with this substrate, the thin films grow in comparable quality including in-plane order as we reported previously for Mn$_2$Au(001) grown on a Ta(001) buffer layer on the same type of substrate \cite{Jou15}. All samples have a thickness of 80~nm and were capped in-situ by $1.8$~nm of Al as a protection against oxidation. The thin films were then patterned by optical lithography and ion beam etching into a cross structure with a central area of $10\times10$~$\rm{\mu m^2}$ to enable current pulsing analogous to our previous experiments, i.\,e.\,with current densities of $\simeq 10^7$~A/cm$^2$ and a pulse durations of 1~ms  \cite{Bod18}.
\begin{figure*}[htb]
	\begin{center}
				\includegraphics[width=1.6\columnwidth, angle=0]{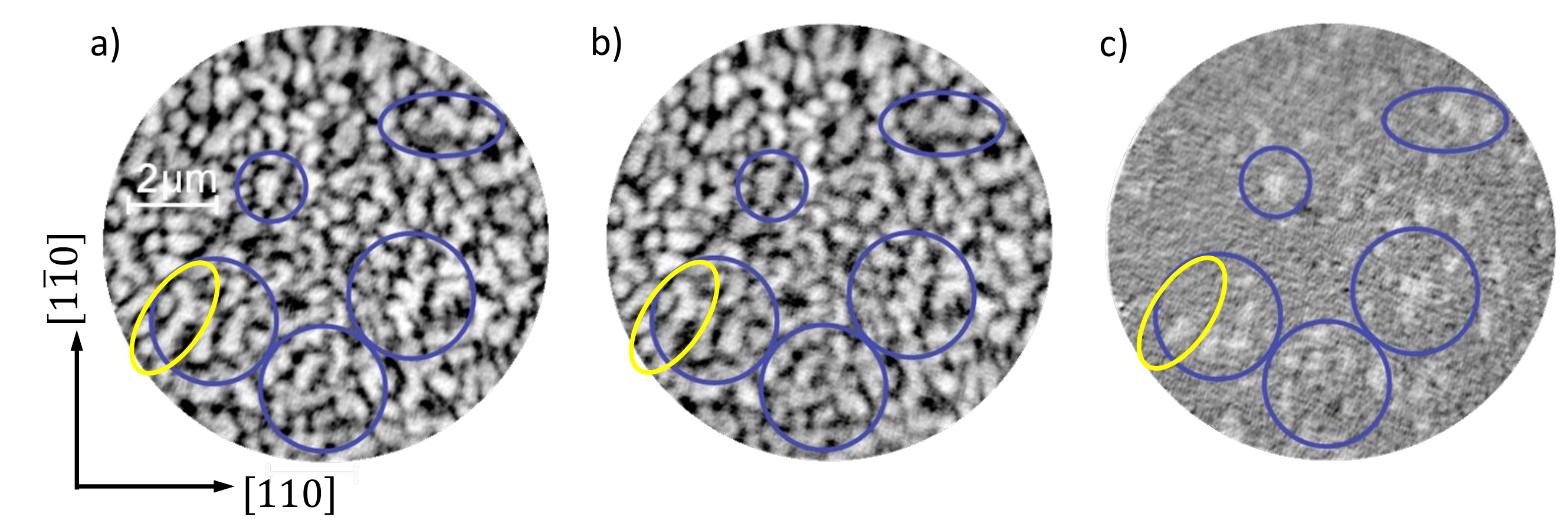}
				\caption{a) XMLD-PEEM asymmetry image of a Mn$_2$Au thin film in the as grown state. b) XMLD-PEEM asymmetry image of the same sample at the same position after the application of a train of current pulses. Both images were obtained with an X-ray direction of incidence parallel [1$\bar1$0], i.\,e.\,with the electric field vector E parallel [110]. c) Difference image of the images a) and b). The blue circles show examples of regions in which the N{\'e}el vector was switched by the current pulses. The yellow circle indicates a grain, which was switched only partially. The field of view has a diameter of 10~$\mu$m and is positioned in the center of a cross structure allowing the injection of current pulses in perpendicular directions.}
				\end{center}
\end{figure*}

The XMLD-PEEM imaging was performed at beamline I06 at Diamond Light Source in the same way as for our previous investigation of the AFM domain pattern of 240~nm thick Mn$_2$Au(001) thin films modified by spin-flop in large external magnetic fields \cite{Sap18}. In this set-up the sample is illuminated under a grazing incidence angle of $16^{\rm o}$ by X-rays linearly polarized in the sample plane. Due to the grazing angle of incidence, the XMLD contrast can barely be generated by a rotation of the polarization direction. Instead, two PEEM images were taken with X-ray energies corresponding to the maximum $E_{MAX}$ and the minimum $E_{MIN}$ of the XMLD as determined in \cite{Sap17}. $E_{MAX}$ roughly corresponds to the ${\rm L_3}$ absorption maximum of Mn, whereas $E_{MIN}$ sits on the low energy side of the edge. However, an asymmetry XMLD-PEEM contrast given by 
\begin{equation}
I_{asymm}=\frac{E_{MAX}-E_{MIN}}{E_{MAX}+E_{MIN}}
\end{equation}
always shows additional to the magnetic contrast generated by the XMLD effect morphological contrast originating from the subtraction two morphology containing images obtained on and off the ${\rm L_3}$ edge. 

Our Mn$_2$Au samples with a thickness of 80~nm used for current induced N{\'e}el vector switching show many more morphological features in PEEM than the 240~nm thick films used in \cite{Sap18}. Fig.\,1~a) shows an asymmetry image of such a sample in the as grown state with the direction of X-ray incidence selected such that magnetic contrast appears for a N{\'e}el vector alignment parallel to the easy [110]- and [1$\bar{1}$0] directions \cite{Sap18, Shi10}.
\begin{figure*}[htb]
	\begin{center}
				\includegraphics[width=2.0\columnwidth, angle=0]{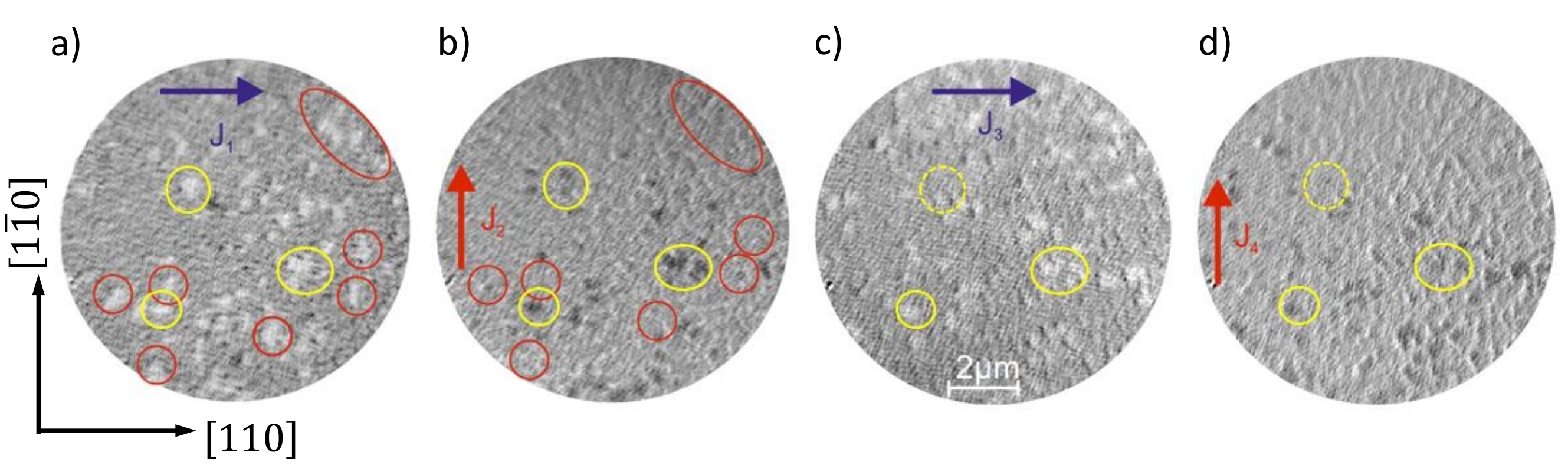}
				\caption{Difference images calculated from XMLD-PEEM asymmetry images obtained before and after the subsequent application of current pulse trains with $I=1.25\times10^7$~A/cm$^2$ along the easy-axis directions [110] and [1$\bar{1}$0]. The current direction for each panel a)-d) is indicated by the blue and red arrows. The yellow circles indicate obvious examples of reversibly switching AFM domains, whereas the red circles indicate examples of domains switching irreversibly.}
				\end{center}
\end{figure*}

Different brightness levels of the morphological grains indicate AFM domains, but  the magnetic contrast is much weaker than the morphological contribution. Correspondingly, upon rotation of a sample by 90$^{\rm o}$ as in \cite{Sap18} we observed only small changes of the XMLD-PEEM images generated by the contrast inversion due to XMLD. However, considering current induced changes of the N{\'e}el vector of the samples, the subtraction of asymmetry XMLD-PEEM images obtained before and after the modification of the magnetic state removes the morphological contrast and results in a difference image clearly showing all modified AFM domains as described below.

We then applied 100 current pulses with an amplitude of $I=1.25\times10^7$~A/cm$^2$ and a length of 1~ms (with a delay of 10~ms between each single pulse) parallel to the [110]-direction and afterward obtained the asymmetry image shown in Fig.\,1~b). It is now possible to observe relative changes of brightness of several morphological grains with respect to the brightness of the surrounding area. The most obvious positions of such changes are indicated by blue circles within the images in Fig.\,1. This is the effect of a N{\'e}el vector reorientation of these grains, which is already visible upon closer inspection of the asymmetry images 1~a) and 1~b). Fig.\,1~c) shows the difference of these images, thereby removing the morphological contrast. Switched grains now appear bright on a homogeneous background. Blue circles at the same position as in 1~a) and 1~b) serve to guide the eye. It is obvious that the applied train of current pulses mostly switched the N{\'e}el vector of complete morphological grains, which indicates that typically the AFM domains correspond to morphological grains and are for these samples of a typical size of $500$~nm. However, in the region indicated by the yellow circles in Fig.\,1 also a partial switching of a grain is observed.
  
Next we study the effects of subsequently applied current pulses along two perpendicular easy directions on the AFM domain structure. After the first set of current pulses already discussed above, a second set of pulses was applied in the perpendicular easy-axis direction, i.\,e.\,along [1$\bar{1}$0] and another XMLD-PEEM asymmetry image was acquired. This procedure was repeated twice always using the same pulse current densities. The corresponding difference images are shown in Fig.\,2~a)-d) with the direction of the current $J$ indicated.
From these images it is obvious that some of the morphological grains/ AFM domains switch reversibly. Examples are indicated by the yellow circles in Fig.\,2. However, other grains, indicated by the red circles, switch irreversibly. This could be explained by local strain induced variations of the magnetocrystalline anisotropy. Alternatively, also an inhomogeneous current density could result in locally irreversible switching behavior. Considering the large AMR effect of several percent associated with the rotation of the N{\'eel} vector \cite{Bod18}, each switching event of a domain results in a redistribution of the inhomogeneous current density, could generate locally irreversible switching behavior.

For a quantitative analysis of the switched fraction of the AFM domains we evaluated the area of the switched domains in the difference images (Fig.\,3, black squares). 
\begin{figure}[htb]
	\begin{center}
				\includegraphics[width=0.9\columnwidth, angle=0]{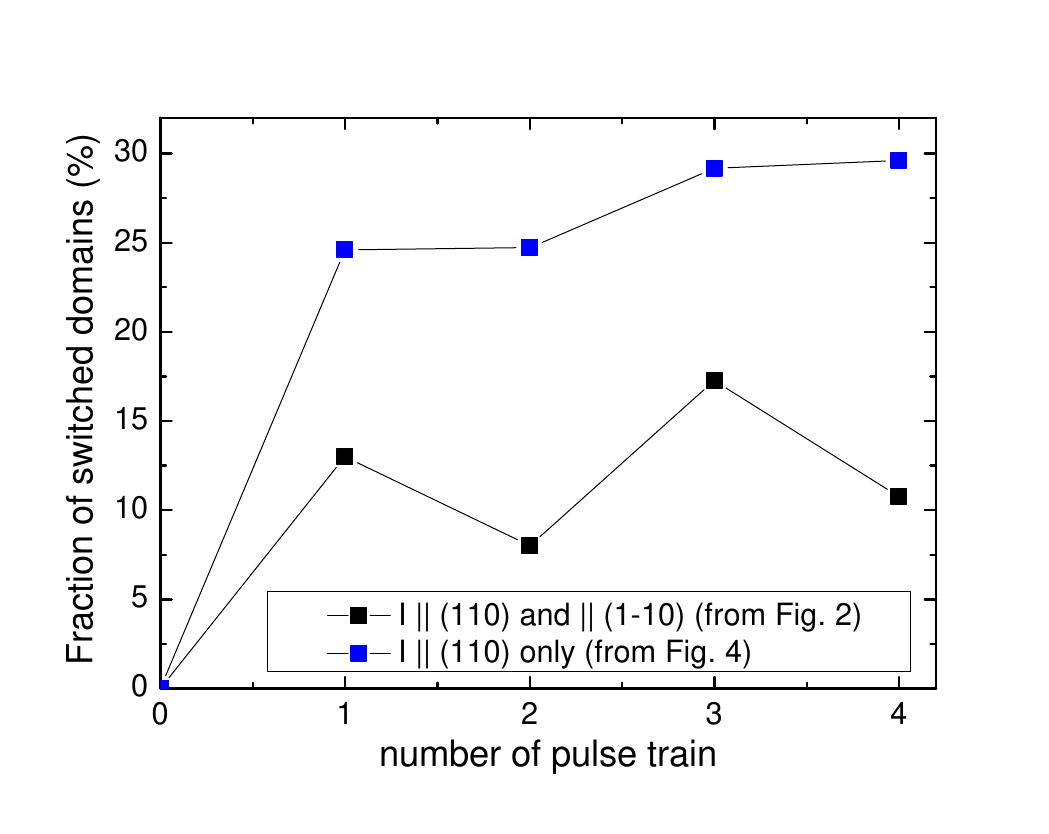}
				\caption{Fraction of switched domains in Figs.\,2 a-d) (black squares), i.\,e.\,for subsequent application of current pulse trains along perpendicular easy directions with constant current densities of $1.25 \times 10^7$~A/cm$^2$ and in Figs. 4 a-d), i.\,e.\,for subsequent application of current pulse trains along the same easy direction with current densities $1.25 \times 10^7$~A/cm$^2$ (pulse trains 1 \& 2) and $1.3\times 10^7$~A/cm$^2$ (pulse trains 3 \& 4)}
				\end{center}
\end{figure}
To obtain the switched fraction the gray scale difference images were converted into black and white images adjusting the threshold in a way that the as determined by eye best correspondence of the switched areas was obtained. The final quantification was obtained from histograms of these binary images. 

Fig.\,3 shows a higher switching probability for the horizontal (pulse trains no.\,1 and 3) than for the vertical (pulse trains no.\,2 and 4) current directions. This is consistent with our observation of a trend toward slightly larger/ smaller resistances when switching the magnetoresistance of Mn$_2$Au by subsequent current pulses in perpendicular directions \cite{Bod18}. The associated breaking of symmetry can be explained by a tilt of the crystallographic (001)-axis with respect to the substrate normal of about 2$^{\rm o}$ as identified by X-ray diffraction, which could affect e.\,g.\,the magnetocrystalline anisotropy.

Next we study the influence of the number of current pulses and of the pulse current density on the resulting fraction of switched AFM domains. For this we investigated a Mn$_2$Au thin film with a prealigned N{\'e}el vector, i.\,e.\,we started from a well defined magnetic configuration with the N{\'e}el vector aligned parallel [1$\bar{1}$0]. This was achieved by exposing the sample to a 60~T external magnetic field pulse (60~ms) at the Helmholtz-Zentrum Dresden-Rossendorf as described in \cite{Sap18} prior to the XMLD-PEEM beam time.

For current densities $\leq 1.2 \times 10^7$~A/cm$^2$ no N{\'e}el vector reorientations were observed. The first set of 100 current pulses with a current density of $1.25 \times 10^7$~A/cm$^2$ along [1$\bar{1}$0] switched already a large fraction of the AFM domains as shown in the difference image obtained from two XMLD-PEEM images acquired before and after the current pulses (Fig.\,4~a).
\begin{figure*}[htb]
	\begin{center}
				\includegraphics[width=2.0\columnwidth, angle=0]{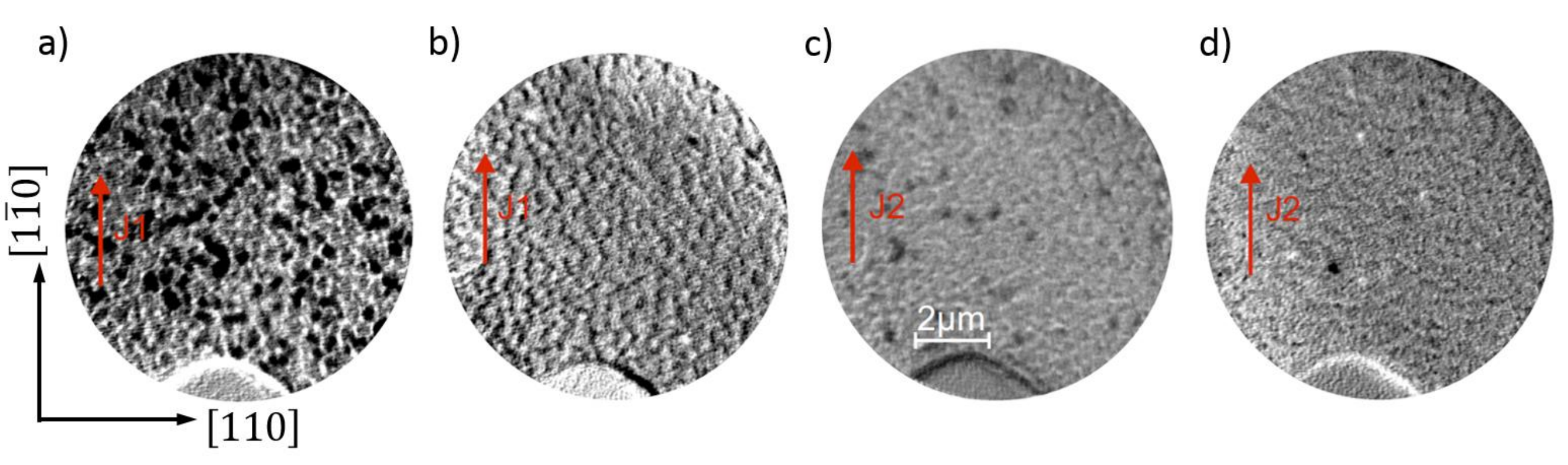}
				\caption{Difference images calculated from XMLD-PEEM asymmetry images obtained before and after the subsequent application of current pulse trains along [1$\bar{1}$0]. The N{\'e}el vector of this sample was prealigned parallel [1$\bar{1}$0] by a spin-flop transition in a 60~T magnetic field $\parallel [110]$ prior to the combined pulsing and XMLD-PEEM experiments. a) after 100 current pulses with $1.25 \times 10^7$~A/cm$^2$, b) after additional 100 current pulses with $1.25 \times 10^7$~A/cm$^2$, c) after additional 100 current pulses with $1.3 \times 10^7$~A/cm$^2$, d) after additional 100 current pulses with $1.3 \times 10^7$~A/cm$^2$.}
				\end{center}
\end{figure*}

Additional 100 pulses with the same current density along the same direction reoriented almost no further domains (Fig.\,4~b). However, another 100 current pulses with an increased current density of $1.3 \times 10^7$~A/cm$^2$ again switched a significant amount of additional domains (Fig.\,4~c). This was again possible (to a smaller extend) by the application of a second set of pulses with the same current density (Fig.\,4~d). At the attempt to continue this procedure with a current density of $1.35\times 10^7$~A/cm$^2$ the sample suffered from electrical breakdown.

Again we quantify the switched fraction of the domains by analyzing the areas of the dark spots in Figs.\,4~a-d). The result is shown in Fig.\,3 (blue squares) and indicates that the first set of 100 current pulses switched about 25~\% of all domains. After all subsequent applications of current pulse trains the total amount of switched domains increased to $\simeq 30$\% before this sample was destroyed.

This relatively large fraction of reoriented domains is consistent with the large resistivity change associated with the current induced N{\'e}el vector manipulation, which we observed previously \cite{Bod18}. Although we could not demonstrate complete switching of all domains, our results together with our previous transport experiments suggest that this can be realized in Mn$_2$Au. Investigating more samples and increasing the pulse current density more carefully, presumably larger switched fractions could be demonstrated. However, this was not possible due to time limitations of beamtime.

In conclusion, we visualized AFM domain patterns of Mn$_2$Au samples, which are reflecting the morphology of these thin films with a typical domain/ grain size of $\simeq 500$~nm. The application of current pulses with a typical current density of $10^7$~A/cm$^2$ resulted mainly in $90^{\rm o}$ rotations of the N{\'e}el vector of domains corresponding to morphological grains. These grains switch partially reversibly and partially irreversibly, which could be explained by local variations of the magnetocrystalline anisotropy or by an inhomogeneous current redistribution. Current induced modifications of the domain pattern were observed exactly within the small window of switching current densities in which we previously observed relatively large resistance changes of the Mn$_2$Au thin films \cite{Bod18}. This represents an important confirmation that our previous assumption of a large AMR effect associated with a reorientation of the N{\'e}el vector in Mn$_2$Au is correct. The fraction of switchable domains increases with increasing current density until destruction of the sample due to current overload. We visualized a maximum switching of $\simeq 30$~\% of all domains in the central area of a patterned cross structure with 10~$\mu$m width, representing a significant fraction of the area previously probed by transport experiments \cite{Bod18}.

{\bf Acknowledgements}

This work is supported by the German Research Foundation (DFG) through the Transregional Collaborative Research Center SFB/TRR173 {\em Spin+X}, Projects A03 and A05. We thank the Diamond Light Source for the allocation of beam time under Proposal No. SI20534-1. We acknowledge the support of the HLD at HZDR, member of the European Magnetic Field Laboratory (EMFL), and by EPSRC (U.K.) via its membership to the EMFL (Grants No. EP/N01085X/1 and No. NS/A000060/1). 

\end{document}